\documentclass[preprint]{interact} 

\usepackage{epstopdf}
\usepackage{subfigure}

\theoremstyle{plain}

\theoremstyle{definition}

\theoremstyle{remark}

\usepackage[latin1]{inputenc}

\usepackage{pdfpages}
\usepackage{braket}
\usepackage[section]{placeins}

\begin{document}



\title{High Visibility in Two-Color Above-Threshold Photoemission from Tungsten Nanotips in a Coherent Control Scheme}

\author{
\name{Timo Paschen\textsuperscript{a}$^{\ast}$\thanks{$^\ast$Corresponding author. Email: timo.paschen@fau.de}, Michael F\"orster\textsuperscript{a,b}, Michael Kr\"uger\textsuperscript{a,b}, Christoph Lemell\textsuperscript{c}, Georg Wachter\textsuperscript{c}, Florian Libisch\textsuperscript{c}, Thomas Madlener\textsuperscript{c}, Joachim Burgd\"orfer\textsuperscript{c} and Peter Hommelhoff\textsuperscript{a,b,d}}
\affil{\textsuperscript{a}Department of Physics, Friedrich-Alexander-Universit\"at Erlangen-N\"urnberg (FAU), Staudtstra\ss{}e 1, 91058 Erlangen, Germany, EU;
\textsuperscript{b}Max Planck Institute of Quantum Optics, Hans-Kopfermann-Stra\ss{}e 1, 85748 Garching, Germany, EU; \textsuperscript{c}Institute for Theoretical Physics, Vienna University of Technology, 1040 Vienna, Austria, EU; \textsuperscript{d}Max Planck Institute for the Science of Light, Staudtstra\ss{}e 2, 91058 Erlangen, Germany, EU}
\received{\today}
}

\maketitle

\begin{abstract}
In this article we present coherent control of above-threshold photoemission from a tungsten nanotip achieving nearly perfect modulation. Depending on the pulse delay between fundamental ($1560\,\mathrm{nm}$) and second harmonic ($780\,\mathrm{nm}$) pulses of a femtosecond fiber laser at the nanotip, electron emission is significantly enhanced or depressed during temporal overlap. Electron emission is studied as a function of pulse delay, optical near-field intensities, DC bias field, and final photoelectron energy. Under optimized conditions modulation amplitudes of the electron emission of 97.5\% are achieved. Experimental observations are discussed in the framework of quantum-pathway interference supported by local density of states (LDOS) simulations.
\end{abstract}


\section{Introduction}
Field enhancement \cite{novotny2012} at nanotips has enabled studies of nonlinear and strong-field physics with moderate laser powers \cite{schenk2010,kruger2011,kruger2012,herink2012,piglosiewicz2014} and confines and enhances electron emission. Recently, the excellent transverse coherence known from DC field emission has been shown to persist in photoemission from sharp nanotips \cite{ehberger2015}. This further highlights the promise of tips as an ultrafast laser-triggered electron source of exceptional beam quality \cite{hoffrogge2014,muller2016,vogelsang2015,feist2015}.

The mixing ratio and the relative phase of two-color laser fields augment the set of parameters to tune and to control dynamics on the (sub-)femtosecond time scale. Using such fields, above-threshold ionization of atoms \cite{muller1990}, high-harmonic generation \cite{watanabe1994}, molecular orientation \cite{de2009} and polarization control of terahertz waves \cite{dai2009} have been investigated.

So far, two-color coherent control studies have mostly been performed with gaseous systems or macroscopic surfaces. Here we present conclusive evidence that the localized emission of nanotips allows us to achieve nearly perfect control of electron emission yield with two-color interference. Control is achieved by variation of the phase between a 1560 nm drive pulse and a weak second harmonic admixture at 780 nm. In-situ inspection of the sample surface aids in obtaining a well-defined electron emitter to surpass the limitations of focal averaging and inhomogeneous broadening. Since the nanotip is much smaller than the focal spot sizes, it singles out local field intensities and phases, so that the inhomogeneous distribution in the laser focus does not play a detrimental role. In this contribution we expand our findings from \cite{foerster2016} and take advantage of the solid-state nature of the tip to investigate the influence of a strong DC bias field on the degree of control achievable in two-color photoemission by the optical phase delay. We find that large DC bias fields ($\sim$ GV/m) inhibit optical control. With optimized parameters we find a contrast of the photocurrent as function of phase delay of 97.5\%.

\section{Experimental setup}

\begin{figure}
\begin{center}
\includegraphics[width=0.5\textwidth]{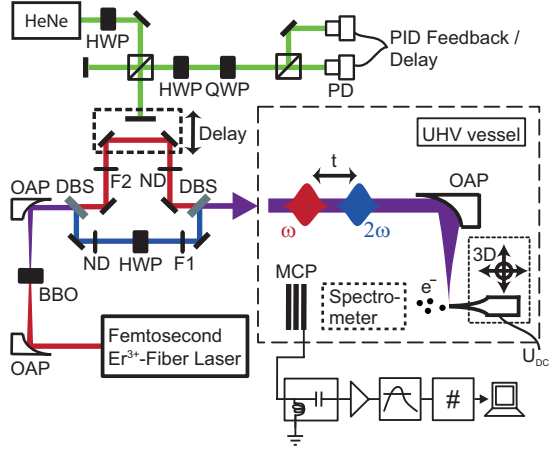}
\caption{Experimental setup. Pulses from an Erbium-doped fiber laser are partially frequency-doubled in a non-linear crystal. The resulting beams are separated in a dichroic Mach-Zehnder-type interferometer to introduce time (and thereby phase) delay. Color and interference filters in both arms ensure a strict separation of fundamental and second harmonic. The delay stage can be locked to an arbitrary position using a second interferometer. Both $\omega$ and $2\omega$ pulses are tightly focused onto the nanotip under UHV conditions by an off-axis parabolic mirror. Emitted electrons from the tip are measured using a MCP detector or a retarding-field spectrometer. Abbreviations are: OAP - off-axis parabolic mirror, BBO - beta-barium borate crystal, DBS - dichroic beam splitter, ND - neutral density filter wheel, HWP - half wave plate, QWP - quarter wave plate, F1/F2 - color and interference filters, PD - photodetector, MCP - microchannel plate detector.}
\label{experimental_setup}
\end{center}
\end{figure}

Our setup is depicted in Fig.\ \ref{experimental_setup}. An amplified Erbium-doped fiber laser system emits laser pulses with a pulse length of 74 fs at a central wavelength of 1560 nm. The pulses are focused into a beta-barium borate (BBO) crystal by an off-axis parabolic mirror (OAP) for second harmonic generation to yield laser pulses with 780 nm central wavelength and a pulse length of 64 fs. Because of the parametric nature of second harmonic generation fundamental and second harmonic pulses are locked in phase. In a Mach-Zehnder-like interferometer fundamental and second harmonic beams are separated and, after manipulation, re-combined with the help of dichroic mirrors. The fundamental pulse can be delayed by virtue of a piezo-driven delay stage. Both interferometer arms contain neutral density filter wheels for intensity control; the second harmonic path is equipped with an additional waveplate for polarization adjustment. A second interferometer with a helium-neon laser is used for delay calibration and stabilization (see Fig.\ \ref{experimental_setup}, green laser path) employing \textit{Pancharatnam's phase} \cite{wehner1997}. For calibration, the differential photodiode signal is recorded by a data acquisition (DaQ) card and used to determine the piezo movement in post-processing. For active stabilization a control voltage is sent to the piezo stage from a proportional-integral (PI) feedback loop using the photodiode signal as input.

After recombination, fundamental and second harmonic pulses are tightly focused onto the nanotip inside a UHV chamber at a base pressure of $5\cdot 10^{-10}$ mbar using an off-axis parabolic mirror. The polarization vectors of fundamental and second harmonic light are parallel to the tip axis. The tip is mounted on a piezo-controlled 3D-translation stage and is biased with a negative voltage, resulting in a static field of up to $|-2\,\mathrm{GV/m}|$ for photoemission experiments. Electron emission is detected either spatially resolved by a micro-channel plate (MCP) detector or energy-resolved using a retarding-field spectrometer. Detected events are discriminated, transformed into TTL pulses, and afterwards counted by a DaQ card. For the presented measurements in this article the Keldysh parameter $\gamma = \sqrt{\phi_{310}/2U_p}$ is for maximum intensities $\gamma_{\omega} \approx 2.7$ and $\gamma_{2\omega} \approx 47$. Here, $\phi_{310}$ is the work function of the (310) plane of tungsten and $U_p$ the ponderomotive energy of the electron in the respective tip-enhanced near field. For the nanotip used in the presented experiment we estimate field enhancement factors of $|\xi_{\omega}|=7$ and $|\xi_{2\omega}|=6$.

The tungsten tips are electrochemically etched from single-crystalline wire oriented along the [310]-direction utilizing the \textit{two-lamellae drop-off} technique \cite{klein1997}. Thereby the low-work-function plane (310) points in the forward direction. The tips typically display a cone-like shank that gradually reduces to a nanometer-sized apex. For further characterization of the surface of the tip apex \textit{Field Ion Microscopy} (FIM) is applied \cite{muller1960} and thereby a radius of curvature of the hemispherical apex of the employed tips of about $10\,\mathrm{nm}$ is determined.


\section{Experimental results}
\begin{figure}
\begin{center}
\includegraphics[width=0.5\textwidth]{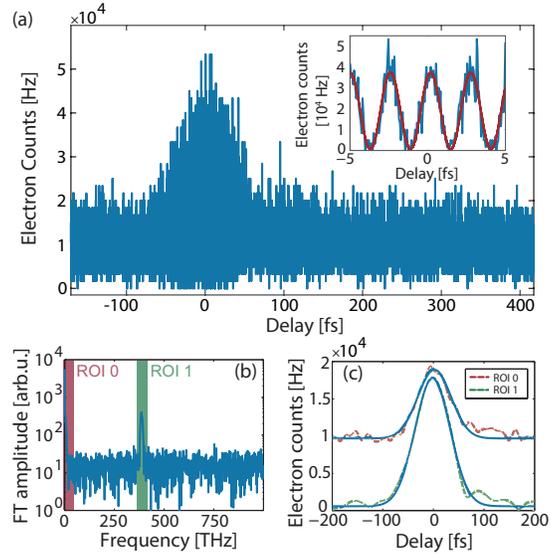}
\caption{Exemplary two-color electron emission and analysis. a) Emitted current from the tip as a function of delay of fundamental and second harmonic pulses with $I_{\omega}=330$ GW/cm$^2$, $I_{2\omega}=6.6$ GW/cm$^2$ (estimated peak intensity at the tip apex) and a static field of $E_{DC}=0.32$ GV/m. Depending on the optical phase a dramatic increase or decrease of emission is noticeable around $t=0$. The inset shows the overlap region in more detail. A high-contrast modulation of the electron count rate is visible. A fit function (red solid line) reveals a sinusoidal behavior and yields a visibility of $94\%$. Under optimized conditions we achieved a contrast of 97.5\% in a different measurement, see Fig.\ \ref{98percent_visbility}. b) Fourier transformation (FT) of the data shown in (a). Two distinct components of the signal can be extracted: A DC-part at low frequencies (highlighted in red) and a peak at 390 THz (highlighted in green). c) Inverse-transformation of the two individual parts ROI 0 and ROI 1 and additional Hilbert transformation of data in ROI 1 (red and green dashed lines) with Gaussian fits to the data (blue solid lines).}
\label{example_2color_signal}
\end{center}
\end{figure}

When we vary the delay $t$ between fundamental and second harmonic pulses we observe a distinct change in electron emission from the tip when the pulses temporally overlap [see Fig.\ \ref{example_2color_signal}(a)]. On top of an overall increase of electron emission the current is modulated with a high contrast even for a very weak admixture of 2\% second harmonic intensity. Compared to the case of non-overlapping pulses the electron emission is in this measurement increased and decreased by a factor of 3.7 and 0.12, respectively. The inset of Fig. \ref{example_2color_signal}(a) demonstrates that the modulation can be well approximated with a sinusoidal function (red solid line).

For quantitative investigation we Fourier-transform the measured data and obtain the dominant frequency components of the phase-dependent electron emission. The Fourier spectrum of the data of Fig.\ \ref{example_2color_signal}(a) is depicted in (b). It is clearly visible that only two main frequency regions contribute, marked ROI 0 (increased average count rate, red area) and ROI 1  at a frequency of 390 THz (amplitude of modulation, green area). The variations of ROI 0 and ROI 1 as function of the delay $t$ are shown in Fig.\ \ref{example_2color_signal}(c). From these data the cooperative contributions in ROI 0 and ROI 1 during temporal overlap are extracted by fitting Gaussian functions of the form
\begin{align}
F_i(t) = A_i+B_i\cdot\exp{(-4\ln{2} t^2/W_i^2)} \label{gauss_fit}
\end{align}
and evaluating the parameters $B_i$. Here, $W_i$ are the full widths at half maximum (FWHM) and $i\in\{0,1\}$.

\begin{figure}
\begin{center}
\includegraphics[width=1\textwidth]{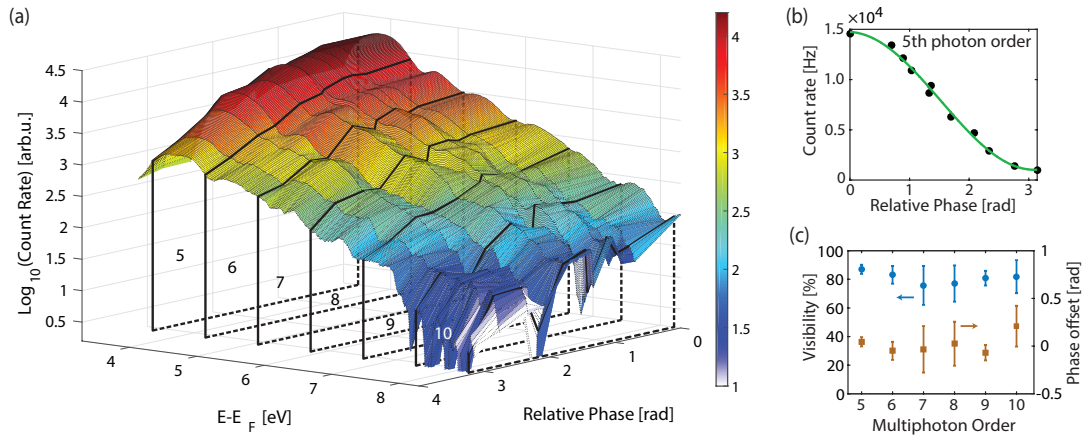}
\caption{a) Electron energy spectra for 11 different phases between $\omega$ and $2\omega$ fields. Near-field intensities are $I_{\omega}=1270$ GW/cm$^2$ and $I_{2\omega}=17$ GW/cm$^2$ and the static field is $E_{DC}=0.37\,\mathrm{GV/m}$. Above-threshold photon orders are indicated with numbers 5 - 10. An exponential drop in count rate towards higher electron energies can be seen for all phases. b) Mean count rate of the 5th multiphoton order as function of $\omega-2\omega$ phase. Green solid line is a sinusoidal fit to the data. c) Visibilities (blue) and phase offsets (brown) for each multiphoton order extracted from individual sinusoidal fits.}
\label{2d_electron_spectra}
\end{center}
\end{figure}

In Fig. \ref{2d_electron_spectra}(a) we show the electron count rate as a function of electron energy and relative phase between the two colors. Multiphoton orders are clearly visible and the count rate drops exponentially with increasing electron energy. By changing the optical phase the emission is homogeneously enhanced or suppressed in a sinusoidal pattern, independent of the final electron energy. For
additional evaluation, the spectra are divided into individual photon orders numbered from 5 to 10. The width of the sections corresponds to the fundamental central photon energy of 0.8 eV. For each relative phase setting the mean count rate of the sections is calculated. In Fig. \ref{2d_electron_spectra}(b) the results are shown exemplarily for the 5th photon order. Sinusoidal fit functions are used to evaluate this data further to obtain the visibility and phase offset with respect to maximum overall current for all multiphoton orders [see Fig. \ref{2d_electron_spectra}(c)]. Each visibility value agrees with the globally observed visibility of 85\% for the shown data set within the error bar. The phase offset values scatter around zero, which indicates equal phases for all multiphoton orders.
In \cite{foerster2016} a more detailed analysis of the data shown in Figs. \ref{example_2color_signal} and \ref{2d_electron_spectra} is presented.

\begin{figure}
\begin{center}
\includegraphics[width=0.5\textwidth]{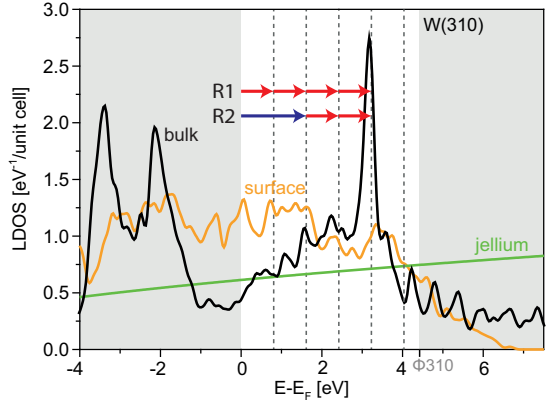}
\caption{Ground-state local density of states (LDOS) simulations for tungsten in (310) direction. Simulated are surface LDOS (orange line) and bulk DOS (black line). For comparison, the DOS for a monotonically increasing jellium is shown in green. A distinct peak at $4\hbar\omega=3.2$ eV in the case of the bulk simulation is visible. Red and blue arrows indicate fundamental and second harmonic photon energies. On display are the two possible paths of the quantum pathway model to the bulk state.}
\label{ldos_simulationen}
\end{center}
\end{figure}

The observed sinusoidal modulation of the photo-emitted current from the nanotip is at variance with a strong-field tunneling model employing the rate of Yudin and Ivanov \cite{yudin2001} which would anticipate an exponential modulation with the amplitude of the combined fields at the tip. Time-dependent density functional theory (TDDFT) calculations for a 1d-jellium model also fail to reproduce our experimental findings. Instead, the monotonically increasing density of states of the jellium supports the strong-field result rather than the experimentally observed sinusoidal behavior. Ground-state density functional theory (DFT) simulations in 3d for tungsten reveal the reason for the discrepancy of experiment and simulation (see Fig.\ \ref{ldos_simulationen}): The local density of states (LDOS) of tungsten is modulated considerably in the energy region between Fermi energy and vacuum level. A pronounced peak in the bulk density of states is visible at $4\hbar\omega$ which acts as a doorway state for further emission: Multiphoton electron emission can be resonantly enhanced compared to the predictions of the jellium model.

The scalings of the parameters $A_i$ and $B_i$ with fundamental and second harmonic intensities, the identical behavior of electrons irrespective of their final kinetic energy and the bulk state at $E_F+4\hbar\omega$ imply an emission scheme where just two pathways contribute to electron emission (see Fig.\ \ref{ldos_simulationen} and \cite{foerster2016}). From the Fermi level at least 4 fundamental photons (red arrows) are needed to reach the doorway state at an energy of $E_F+4\hbar\omega$. This pathway involving only fundamental photons is labeled R1. Another possible path to reach this state is shown as pathway R2. Here two fundamental photons and one additional second harmonic photon are required for the electron to be excited. Both pathways reach the same state, enabling interference due to the coherent nature of the excitation. From the excited state emission proceeds further. Emission channels incorporating more than one second harmonic photon can be neglected at small $2\omega$ admixtures due to their very low probability.

\begin{figure}
\begin{center}
\includegraphics[width=0.5\textwidth]{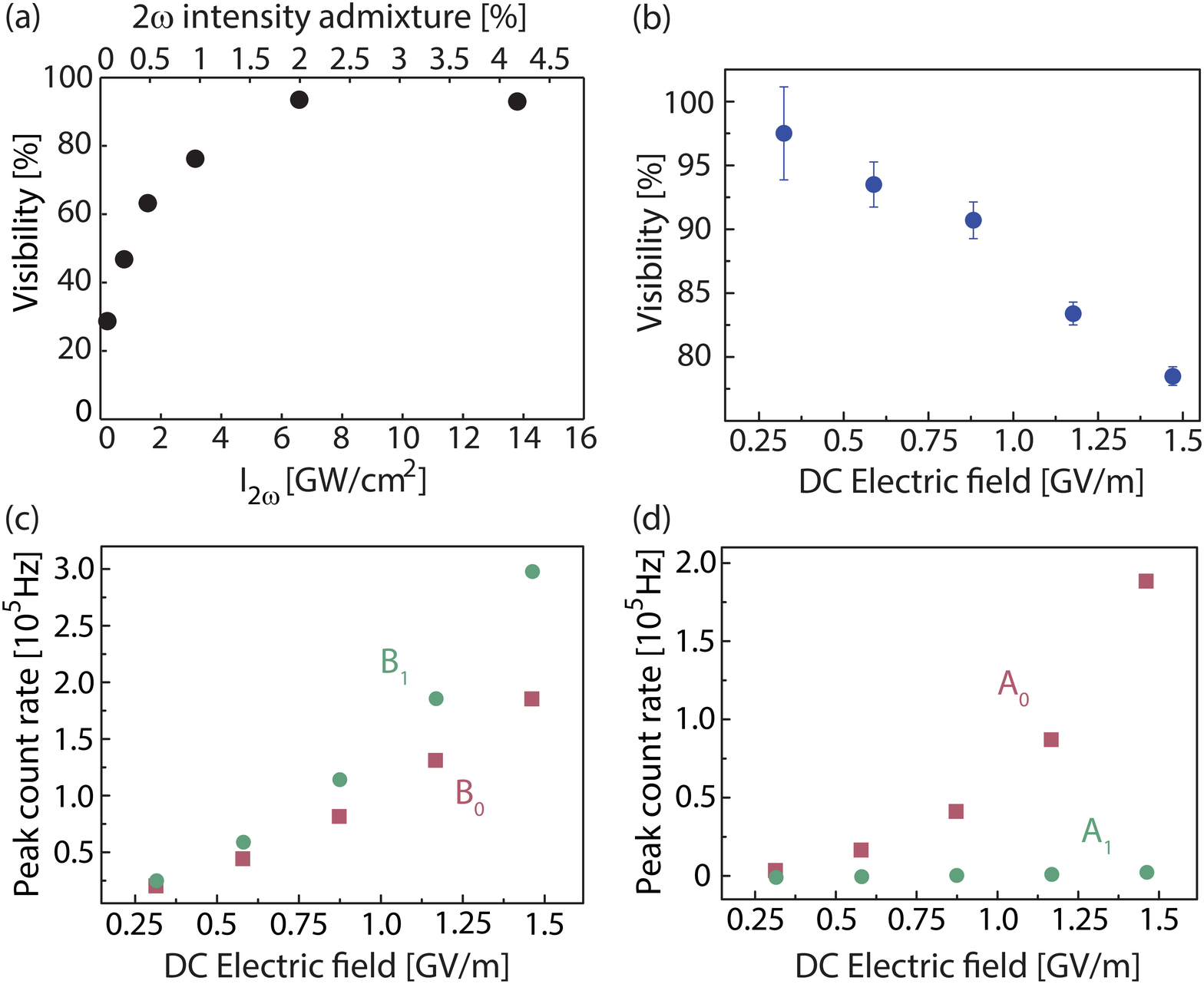}
\caption{Visibility of  emitted current as function of control parameters. a) Visibility as a function of second harmonic intensity. For $I_\omega=330$ GW/cm$^2$ and a static field of $E_{DC}=0.32$ GV/m the visibility is maximum at an admixture of 2\% second harmonic intensity (black circles). b) Visibility dependence on bias tip voltage. Maximum visibility is reached for lowest tip voltage. A monotonic trend is evident. c) and d) Scaling of the peak amplitudes $B_i$ and signal offsets $A_i$ according to Eq. \ref{gauss_fit} with tip bias voltage. $A_0$ and $B_0$ are depicted as red squares, $A_1$ and $B_1$ as green circles. For the data shown in (b) - (d) near-field intensities are $I_\omega=250$ GW/cm$^2$ and $I_{2\omega}=12.9$ GW/cm$^2$.}
\label{scaling_visbility}
\end{center}
\end{figure}
The visibility of the modulated electron count rate is given by
\begin{align}
V=\frac{N_{max}-N_{min}}{N_{max}+N_{min}} \hspace{0.5cm}.
\end{align}
Here $N_{max}$ and $N_{min}$ are maximum and minimum count rate determined in temporal overlap with the help of sinusoidal fits. In Fig.\ \ref{scaling_visbility}(a) the visibility values for a data set with varying $I_{2\omega}$-admixture, including the data of Fig.\ \ref{example_2color_signal}, are shown. In this case, a 2\% admixture of second harmonic leads to the highest visibility. For the data point with the highest admixture a new frequency component at $4\omega$ is visible in the Fourier spectrum, which indicates the onset of higher-order multiphoton processes not covered by the two-pathway model.

A further control knob for the visibility of the emission-current modulation is the tip bias voltage [Fig.\ \ref{scaling_visbility}(b)-(d)]. In addition to optical fields, also strong DC fields of up to $|-2\,\mathrm{GV/m}|$ can be applied to the nanotip apex for a photoemission experiment, limited by the onset of DC field emission. Thus, for the small second harmonic intensities used in our experiment, the bias field is comparable to the maximum field of the second harmonic.

For all fundamental intensities and second harmonic admixtures we observe a monotonic decrease of the evaluated visibility with increasing magnitude of the bias voltage. This occurs despite stronger cooperative signals signified by larger fit parameters $B_0$ and $B_1$ [Fig.\ \ref{scaling_visbility}(c)]. The decrease in visibility is mainly caused by the growth of the non-cooperative contribution to electron emission [$A_0$, see Fig.\ \ref{scaling_visbility}(d)], which grows faster with the bias field $E_{DC}$ than $B_0$ and $B_1$. We find $A_0$ to be almost identical to and to show the same dependence on $E_{DC}$ as count rates obtained by fundamental color pulses only, possibly due to increased photo-assisted field emission after one or two photon capture (note the increased surface LDOS in Fig.\ \ref{ldos_simulationen} at $\hbar\omega$ and 2$\hbar\omega$) \cite{hommelhoff2006}. Thus, for maximum two-color phase-control of the emitted current the single-color background emission should be reduced to a minimum by decreasing the bias field applied to the tip.

\begin{figure}
\begin{center}
\includegraphics[width=0.5\textwidth]{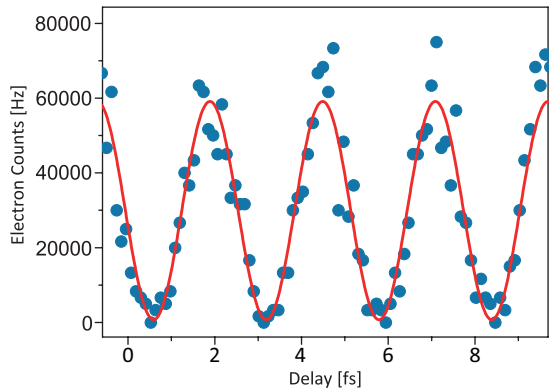}
\caption{Highest achieved visibility of 97.5\% in pulse overlap with intensities $I_{\omega}=250$ GW/cm$^2$, $I_{2\omega}=12.9$ GW/cm$^2$ and a static field of $E_{DC}=0.32$ GV/m. Blue dots are measured data, the red solid line is a sinusoidal fit to the data to determine the visibility. Time $t = 0$ is set arbitrarily.}
\label{98percent_visbility}
\end{center}
\end{figure}

With optimized control parameters ($I_\omega=250\,\mathrm{GW/cm}^2$, $I_{2\omega}=12.9\,\mathrm{GW/cm}^2$, and $E_{DC}=0.32\,\mathrm{GV/m}$) we could reach a maximum visibility of 97.5\% for two-color electron emission. In Fig.\ \ref{98percent_visbility} a close-up of the modulated electron emission in the region of temporal overlap of the two colors for this maximum visibility is presented. The minimum count rate drops nearly to zero resulting in almost completely suppressed electron emission from the nanotip. Note that, while we observe that the control over the emitted current with phase delay persists and is even strengthened, for the high second harmonic intensity admixture of 5\% for the measurements presented in Figs.\ \ref{scaling_visbility} and \ref{98percent_visbility} the simple two-pathway model discussed above is not sufficient and has to be amended by additional channels.

\section{Outlook and Conclusion}
In this article we demonstrated coherently controlled two-color above-threshold photoemission from a metal nanotip. Optimizing fundamental and second harmonic intensities, as well as the static field at the tip, an optimized visibility of 97.5\% was obtained, which is amongst the highest values for two-color experiments \cite{ackermann2014}. Given the solid-state nature of the nanotip the degree of control of the total electron emission is surprisingly high. A quantum pathway interference model is capable of describing the sinusoidal behavior of electron emission as function of the optical phase between the two colors for small $I_{2\omega}$ admixtures. This model is further supported by DFT calculations that indicate resonant electron emission enhancement using a doorway state at $4\hbar\omega$.

Nanotips may in the future be used as a nanometric probe for light phases \cite{chen1990}. Furthermore, utilizing polarization-shaped two-color laser pulses \cite{brixner2001} in conjunction with the polarization-sensitive near field, and the spatially varying work function at the nanotip surface \cite{yanagisawa2009} will allow the generation of femtosecond user-defined electron bunches.
\section*{Funding}
This project was funded in part by the ERC grant `Near Field Atto,' by DFG SPP 1840 `QUTIF,' and by the Austrian Science Fund (FWF) within the special research projects SFB-041 `ViCoM' and SFB-049 `Next Lite' and project P21141-N16. M.F. and G.W. acknowledge support by the IMPRS-APS.

\bibliography{literatur_modern_optics}

\begin{thebibliography}{25}
\providecommand{\natexlab}[1]{#1}
\providecommand{\noopsort}[1]{}
\providecommand{\printfirst}[2]{#1}
\providecommand{\singleletter}[1]{#1}
\providecommand{\switchargs}[2]{#2#1}

\bibitem[1]{novotny2012}
Novotny, L.; Hecht, B. Principles of nano-optics.  Cambridge University Press:
  Cambridge, 2012.

\bibitem[2]{schenk2010}
Schenk, M.; Kr{\"u}ger, M.; Hommelhoff, P. {\em Physical Review Letters}  {\bf
  2010}, {\em 105} (25), 257601.

\bibitem[3]{kruger2011}
Kr{\"u}ger, M.; Schenk, M.; Hommelhoff, P. {\em Nature}  {\bf 2011}, {\em 475}
  (7354), 78--81.

\bibitem[4]{kruger2012}
Kr{\"u}ger, M.; Schenk, M.; Hommelhoff, P.; Wachter, G.; Lemell, C.; et~al.
  {\em New Journal of Physics}  {\bf 2012}, {\em 14} (8), 085019.

\bibitem[5]{herink2012}
Herink, G.; Solli, D.; Gulde, M.; et~al. {\em Nature}  {\bf 2012}, {\em 483}
  (7388), 190--193.

\bibitem[6]{piglosiewicz2014}
Piglosiewicz, B.; Schmidt, S.; Park, D.J.; Vogelsang, J.; Gro{\ss}, P.;
  Manzoni, C.; Farinello, P.; Cerullo, G.; et~al. {\em Nature Photonics}  {\bf
  2014}, {\em 8} (1), 37--42.

\bibitem[7]{ehberger2015}
Ehberger, D.; Hammer, J.; Eisele, M.; Kr{\"u}ger, M.; Noe, J.; H{\"o}gele, A.;
  et~al. {\em Physical Review Letters}  {\bf 2015}, {\em 114} (22), 227601.

\bibitem[8]{hoffrogge2014}
Hoffrogge, J.; Stein, J.P.; Kr{\"u}ger, M.; F{\"o}rster, M.; Hammer, J.;
  Ehberger, D.; Baum, P.; et~al. {\em Journal of Applied Physics}  {\bf 2014},
  {\em 115} (9), 094506.

\bibitem[9]{muller2016}
Mueller, M.; Kravtsov, V.; Paarmann, A.; Raschke, M.B.; et~al. {\em ACS
  Photonics}  {\bf 2016}, {\em 3} (4), 611--619.

\bibitem[10]{vogelsang2015}
Vogelsang, J.; Robin, J.; Nagy, B.J.; Dombi, P.; Rosenkranz, D.; Schiek, M.;
  Gro{\ss}, P.; et~al. {\em Nano Letters}  {\bf 2015}, {\em 15} (7),
  4685--4691.

\bibitem[11]{feist2015}
Feist, A.; Echternkamp, K.E.; Schauss, J.; Yalunin, S.V.; Sascha, S.; et~al.
  {\em Nature}  {\bf 2015}, {\em 521} (7551), 200--203.

\bibitem[12]{muller1990}
Muller, H.; Bucksbaum, P.; Schumacher, D.; et~al. {\em Journal of Physics B:
  Atomic, Molecular and Optical Physics}  {\bf 1990}, {\em 23} (16), 2761.

\bibitem[13]{watanabe1994}
Watanabe, S.; Kondo, K.; Nabekawa, Y.; Sagisaka, A.; et~al. {\em Physical
  Review Letters}  {\bf 1994}, {\em 73} (20), 2692.

\bibitem[14]{de2009}
De, S.; Znakovskaya, I.; Ray, D.; Anis, F.; Johnson, N.G.; Bocharova, I.;
  Magrakvelidze, M.; Esry, B.; Cocke, C.; Litvinyuk, I. {\em et~al.} {\em
  Physical Review Letters}  {\bf 2009}, {\em 103} (15), 153002.

\bibitem[15]{dai2009}
Dai, J.; Karpowicz, N.; Zhang, X.C. {\em Physical Review Letters}  {\bf 2009},
  {\em 103} (2), 023001.

\bibitem[16]{foerster2016}
F{\"o}rster, M.; Timo, P.; Michael, K.; Lemell, C.; Wachter, G.; Florian, L.;
  Madlener, T.; Burgd{\"o}rfer, J.; et~al. {\em arXiv}  {\bf 2016}, {\em
  1603.01516}.

\bibitem[17]{wehner1997}
Wehner, M.; Ulm, M.; Wegener, M. {\em Optics Letters}  {\bf 1997}, {\em 22}
  (19), 1455--1457.

\bibitem[18]{klein1997}
Klein, M.; Schwitzgebel, G. {\em Review of Scientific Instruments}  {\bf 1997},
  {\em 68} (8), 3099--3103.

\bibitem[19]{muller1960}
M{\"u}ller, E.W. {\em Advances in Electronics and Electron Physics}  {\bf
  1960}, {\em 13}, 83--179.

\bibitem[20]{yudin2001}
Yudin, G.L.; Ivanov, M.Y. {\em Physical Review A}  {\bf 2001}, {\em 64} (1),
  013409.

\bibitem[21]{hommelhoff2006}
Hommelhoff, P.; Sortais, Y.; Aghajani-Talesh, A.; et~al. {\em Physical review
  letters}  {\bf 2006}, {\em 96} (7), 077401.

\bibitem[22]{ackermann2014}
Ackermann, P.; Scharf, A.; Halfmann, T. {\em Physical Review A}  {\bf 2014},
  {\em 89} (6), 063804.

\bibitem[23]{chen1990}
Chen, C.; Elliott, D. {\em Physical Review Letters}  {\bf 1990}, {\em 65} (14),
  1737.

\bibitem[24]{brixner2001}
Brixner, T.; Gerber, G. {\em Optics Letters}  {\bf 2001}, {\em 26} (8),
  557--559.

\bibitem[25]{yanagisawa2009}
Yanagisawa, H.; Hafner, C.; Don{\'a}, P.; Kl{\"o}ckner, M.; Leuenberger, D.;
  Greber, T.; Hengsberger, M.; et~al. {\em Physical review letters}  {\bf
  2009}, {\em 103} (25), 257603.

\end{thebibliography}
\bibliographystyle{tMOP}

\end{document}